# STUDY ON THE FORMATION OF NANO $\tau_3$ PHASE BY MECHANICAL ALLOYING

***T.P. Yadav[1], N.K. Mukhopadhyay[2*], R.S. Tiwari [1] and O.N. Srivastava [1]***

[1]Department of Physics, Banaras Hindu University, Varanasi-221 005 (INDIA)

[2]Department of Metallurgical Engg., I.T., Banaras Hindu University, Varanasi-221 005 (INDIA)

## ABSTRACT

In the present investigation the pure elemental powder mixture of Al (70 at %) Ni (15 at %), Cu (15 at %) was mechanically milled in an attritor ball mill for 10, 20, 40, 60, 80 and 100 hours in hexane medium at 400 rpm. The isothermal annealing of 100 h mechanically milled powder has been done at 700 $^0$C for 20, 40 and 60 hours. The mechanically alloyed powders are characterized using X-ray diffraction, differential thermal analysis and transmission electron microscopy techniques. It was observed that mechanical alloying led to the formation of nano $\tau_3$ phases in $Al_{70}$ $Cu_{15}Ni_{15}$ after 80 h of milling. In the case of 100 h MM and subsequent annealing at 700C for 20, 40 and 60 h, powder exhibited the formation of $\tau_3$ phase with bigger grain sizes. The phase formation and transformations in the above systems have been discussed based.

* E-mail: mukho_nk@rediffmail.com

**INTRODUCTION**

Non-equilibrium processing techniques such as mechanical milling / alloying (MM/MA), rapid solidification etc are being used recently to design the materials with structures desirable for technological applications [1-3]. Particularly, MM/MA process has attracted the attention because of its potential in nanoscience and technology. It seems to provide the route for synthesizing bulk nanocrystalline and amorphous materials from alloys, immiscible systems and intermetallic compounds [4-5].

Ternary Al based alloys have been extensively studied recently for developing as high temperature materials for industrial application [6]. In addition to the complex crystalline phases such as Laves phase, topologically closed phase, vacancy ordered phases; the quasicrystalline phases also appear in many systems. There are efforts to understand the origin of these complex metallic phases and their role for developing the advanced material. It is found that many of the complex crystalline phases are related to quasicrystalline phases. For example, it has been shown that the vacancy ordered phases, designated as $\tau$ phase can be considered as approximant phase to QC structures. The vacancy ordered phases can be described in terms of basic B2 (CsCl) type of cell with vacancy ordering along the [111] direction. The B2 (CsCl) suporstructure is a common structure type in intermetallics, its unit cell contains two different atoms located respectively on the vertex and the centre of a cube. The series of stable vacancy ordered phases (VOPs) and its relation with one dimensional quasiperiodicity has been demonstrated (Chattopadhyay etal.)[7].The different $\tau$ phases can be identified on the basis of the number of divisions made by Bragg peaks along [111] direction in the diffraction patterns. For example $\tau_3$ represents three times ordering along [111] direction of the basic CsCl unit cell and it is exhibited by three equal spacing in the diffraction pattern along [111] direction. VOPs in Al-TM (TM transition metal) system are a special class of structures wherein vacancies in the TM sublattice are ordered on the (111) planes [8]. Vacancy ordered phases ($\tau_3$) in the Al-Cu-Ni system have an arrangements of vacant and filled sites in the truncated Fibonacci sequence along the [111] direction. These phases can be classified in terms of different sequence of ordering among Al atoms, Cu / Ni atoms and vacancies along triad axis of the basic CsCl structure [9].

The aim of the present investigation was to synthesize nano vacancy ordered phase (VOP) by mechanical alloying of Al (70 at %) Cu (15 at %) and Ni (15 at %) and to understand the microstructural and phase stability during subsequent annealing.

## EXPERIMENTAL DETAILS

### Synthesis

Powder mixture of 10 gm Al, Cu, Ni containing 70 at %, Al 15 at %, Cu and 15 at 15 % Ni with Al 400 mesh, Cu 200 mesh, Ni 200 mesh were taken. These elemental powders were mechanically milled in Szegvari Attritor Ball mill with ball to powder ratio of 80 : 1 (with total mass of balls 800 gm), with a revolving speed of 400 rpm. The attritor has a cylindrical stainless steel tank of innerdiameter 13 cm. Hardened steel ball of diameter 6 mm were used. The milling operation was conducted for 0 to 100 hours in hexane medium. Isothermal heat treatment of 100 hours ball milled sample was carried out at $600^0$C for 10, 20 and 40 hours.

### Structural Characterization

The mechanically milled and annealed sample were subjected to structural characterization employing powder X ray diffraction (XRD) Philips PW 1710 diffractometer with $CuK_\alpha$ radiation $\lambda$ = 1.5418A. The experimental conditions such as rating of X ray generator (30 KV, 20 mA) and other diffractometer parameters such as scanning speed, were kept constant for all diffraction experiments performed on different samples, Differential thermal analysis (DTA) with heating rate is 10 °C /min.

Sample thus prepared were studied by TEM using a Philips CM-12 electron microscope employing imaging and diffraction modes.

The grain size and the lattice strain of the sample can be calculated from the integral width of the physical broadening profile. Couchy and Gaussian components can be obtained from the ratio of full width at half maximum intensity ($2\omega$) and integral breadth ($\beta$) [10]. In a single line analysis the apparent crystallite size '*D*' and strain '*e*' can be related to Couchy ($\beta$c) and Gaussian ($\beta$G) widths of the diffraction peak at the Bragg angle $\theta$;

$$D = k\lambda / \beta c \cos \theta \qquad \text{(i)}$$

and

$$e = \beta G / 4 \tan \theta \qquad \text{(ii)}$$

The constituent Couchy and Gaussian components can be given as

$$\beta_c = (a_0 + a_1\psi + a_2\psi 2)\beta$$

$$\beta_G = _b_0 + b_{1/2}_\psi .2\pi_{1/2} + b_1\psi + b_2\psi 2_\beta$$

where $a_0$, $a_1$ and $a_2$ are Couchy constants, $b_0$, $b_{1/2}$, $b_1$ and $b_2$ are Gassian constants and $\psi = 2\omega/\beta$ where $\beta$ is the integral breadth obtained from XRD peak. The values of Couchy and Gaussian constant have been taken from the table of Langford [10]

$a_0$ = 2.0207, $a_1$ = .0.4803, $a_2$ = .1.7756; $b_0$ = 0.6420, $b_{1/2}$ = 1.4187, $b_1$ = .2.2043, $b_2$ = 1.8706

## RESULT AND DISCUSSIONS

Figure 1 shows the XRD patterns of elemental Al, Cu, Ni peaks as a function of milling time, leading to the gradual intermixing and evolution of $\tau_3$ phase. The un-milled sample (fig. 1(a) ) shows the peaks of Al ,Cu and Ni. During milling the peaks of the elemental powder shift towards the position corresponding to the final phases. Finally, after 100 h of mechanical alloying all the pure metal peaks disappeared and broad peaks corresponding to $\tau_3$ phase appeared in fig 1(g). At a milling intensity of 60 h the rate of phase formation was found to be nearly complete and finally it was complete after 100 h of milling. It is clearly seen that, in course of milling, the peak intensities corresponding to elemental powder are decreased and the peaks become broadened, suggesting that a large amount of defects were introduced into the samples. The relative peak intensities of the milled samples are found to be close to the standard values of the $\tau_3$ phase (JCPDS cards: [11]). There were many reflections were absent compared to the standard data. This can be attributed to the fact that the milled powder contains highly disordered phase and as a result the weak intensities in the ordered structures cannot be observed due to very low intensity or nearly zero intensity. By analyzing the peak broadening, the grain size and lattice strain have been determined. The calculated the crystallite size is found in this case ~ 12 nm and the lattice strain ~ 0.675 of the 100h MM of $Al_{70}Cu_{15}Ni_{15}$ powders.

Fig. 2 shows the result of DTA investigation of 100h MM $Al_{70}Cu_{15}Ni_{15}$ powder. In this case, the heating rate is 10 $^{o}$C/min. Some broad hump in the DTA curve are seen which may correspond to annealing out of defects in the system and thereby the energy is released. Furthermore the ordering in the structure can be discerned. However, no sharp exothermic/endothermic peaks were seen which can be related to order/disorder transformation.

Figures 3 (a), (b), (c), show XRD patterns obtained from 100h MM after annealing at 700 $^{o}$C for 20h-60h respectively. In all of the cases the peak broadening is found to be reduced compared with that in the as-milled condition. This can be attributed to strain relaxation along with slight domain coarsening. Fig. 2(c) have been indexed using $\tau_3$ structure with a=8.7 Å which is the superstructure of B2 phase. The formation of disordered $\tau_3$ phase due to mechanical milling has been observed in Al-Cu-Ni alloys system in the present study. It is known that the high energy ball milling technique introduces defects, disordering and thereby destabilize the ordered phase. However in the present case, it is interesting to note the vacancy ordered phase has been directly from MA. The role of heat treatment is to eliminate the defects, induce ordering and increase the grain size of the milled product.

This formation of $\tau_3$ phases was also monitored through rigorous transmission electron microscopic investigation by obtaining selected area electron diffraction patterns and microstructural features at different stages. The initial milling (10 h- 20h) of the starting powder mixture produces a lamellar (or layered) microstructure. Further plastic deformation of these lamellae and interdiffusion of atoms between lamellae took place with increasing milling time. This lamellae microstructure gradually disappeared and a nano-crystalline microstructure has been formed, as shown in fig.4(a-b) . The bright and dark field (figure 4 a-b) TEM image clearly shows fine grains having sizes ranging from 10 to 20 nm. The SAD pattern in figure 4c was indexed by $\tau_3$ (VOP) phase. It is interesting to point out that the $\tau_3$ phase has been designated as an approximant phase of decagonal quasicrystal by Dong et al. [12] derived from B2 structure. It is also interesting to note that the B2 phase has not formed as the transient phase during milling, though there is an intimate relation between B2 phase. The $\tau_3$ phase is characterized by vacancy ordering in a B2 (CsCl) lattice. The vacancy sites are ordered along [111] direction of B2 pseudocell dividing it into three periods, eventually the transformed cell is distorted rhombohedral. This might be due to the fact that in Al-Cu binary neither bcc or B2 phase is not stable which is due to different elctronic effect of Cu compared to the transition metal elements. Moreover the stoichiometry of the present phase is close to the $\tau_3$ phase ($Al_3Ni_2$ type). In fact in the present alloy is also slightly away from the ideal composition of the perfect $\tau_3$, indicating the excess of Al. This can be accommodated in the form of defects or minor loss of Al due to oxidation, which is unavoidable during milling. This also explains why there should be an excess amount of Al in order to obtain the particular composition considering the minor loss. However, in the present milling experiments $\tau_3$ phase was found to be a dominant phase from the beginning. Figs. 5 (a, b) shows the microstructures and diffraction pattern corresponding to more ordered $\tau$3 phase. It is clear that from the present investigation we could synthesize initially the disordered nano $\tau_3$ phase and subsequently ordered $\tau_3$ phase during annealing. This diffraction does not show any kind of disordering as there is no diffuse scattering and it corresponds to the $\tau_3$ phase. The grain coarsening can be seen which has grown from nano sizes to sub micron sizes. The studies are under way to find out the suitable temperature so that the grain coarsening can be limited to nano scale but the perfect ordering can be attained. It will be also interesting to establish the order disorder transformation temperature as well as the ordering of vanacy or Al/Ni/Cu atoms.

Because it appears that this ordering could be related to the disordering among the metallic atoms and not among the vacancy and the metallic atoms. Otherwise we would have got other kind of disordered structures. But in the milled powder we have obtained $\tau_3$ phase, which indicates that the vacancy is somewhat already ordered in the structures.

## Conclusions:

The formation of nano $\tau 3$ vacancy ordered phase in the alloy system, Al-Ni-Cu has been established by mechanically alloying the elemental powder. The nanograin size has been estimated to be around 10-20nm and the strain as 0.675. There is a considerable amount of disordering in the as milled phase and as a result of that many reflections are not visible. The observed reflections are also broadened due to disordering, size and strain effect. However, after annealing at 700 $^{o}$C the more ordering can be seen as the peaks are sharp and many more reflections are present. The crystallite size was found to increase and the strain has completely vanished. The formation of disordered nano $\tau 3$ phase appears to be related to the defects originated during the milling.

**Acknowledgement:**

The authors would like to thank Prof. G.V.S. Sastry, Dr. R.K. Mandal and Dr. M.A.Shaz for many stimulating discussions. The financial support from Ministry of Non-Conventional Energy Sources, New Delhi, India is gratefully acknowledged. One of the authors (T.P.Y) acknowledges to CSIR for a Senior Research Fellowship, during which period the work was completed.

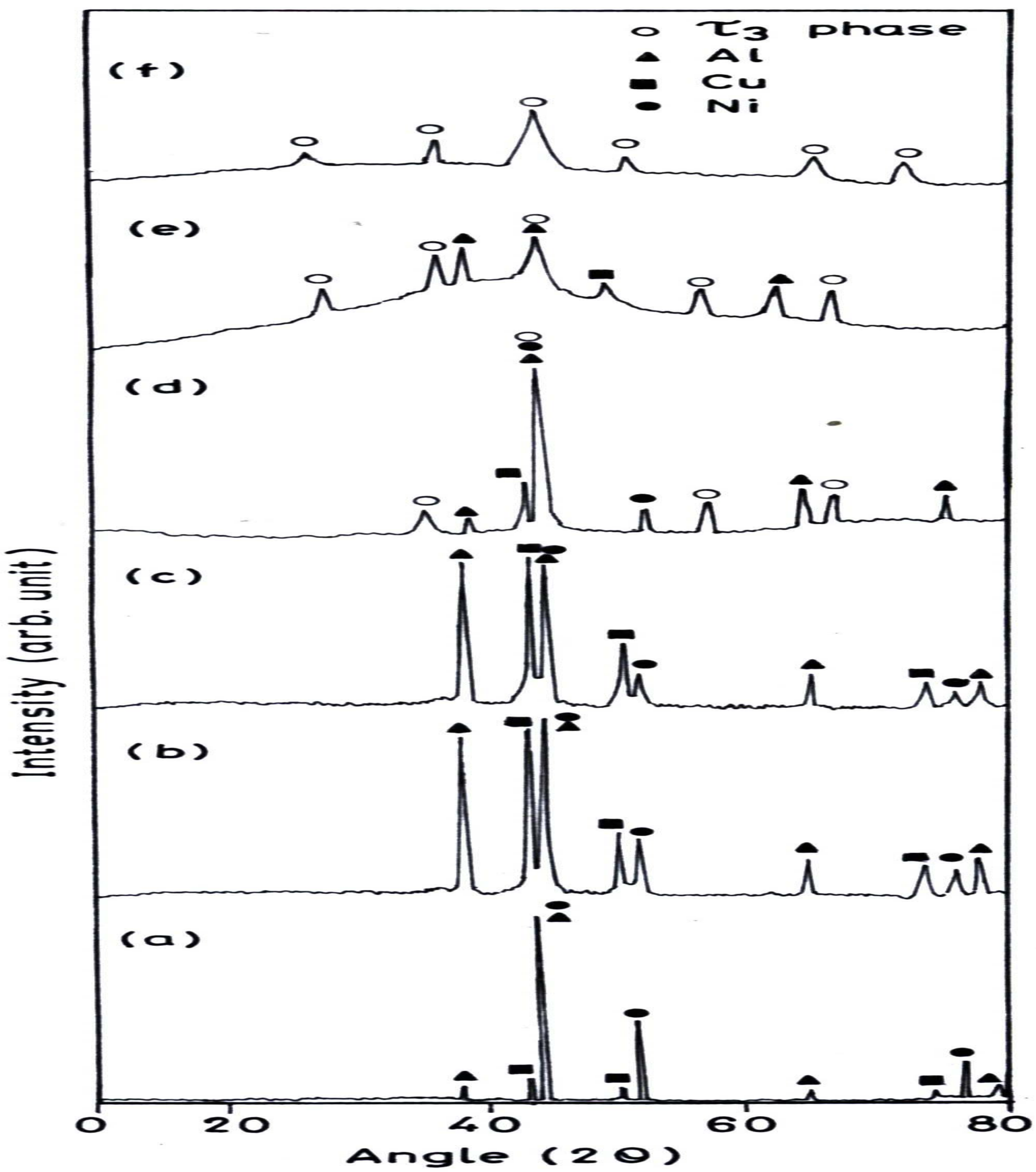

Fig.1: X-ray diffraction patterns for Al70Ni15Cu15 elemental powders (a) as well as milled powders for different milling durations(b-f) ,demonstrating the formation of $\tau$**3** phase.

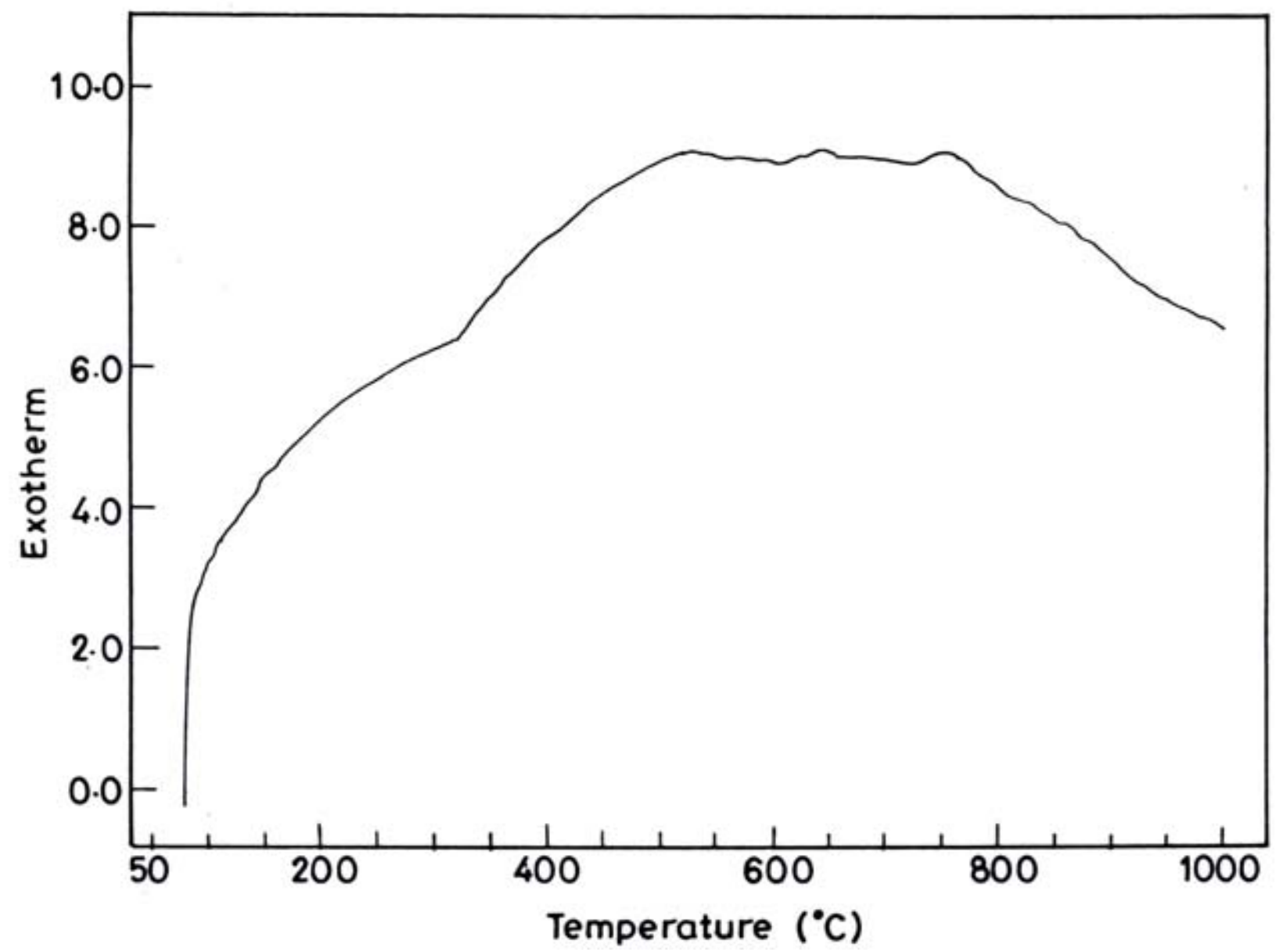

Fig.2: Differential thermal analysis (DTA) curve of the 100 h MM Al-Cu-Ni powders. The trace appears quite complex and indicates a bored exothermic peak.

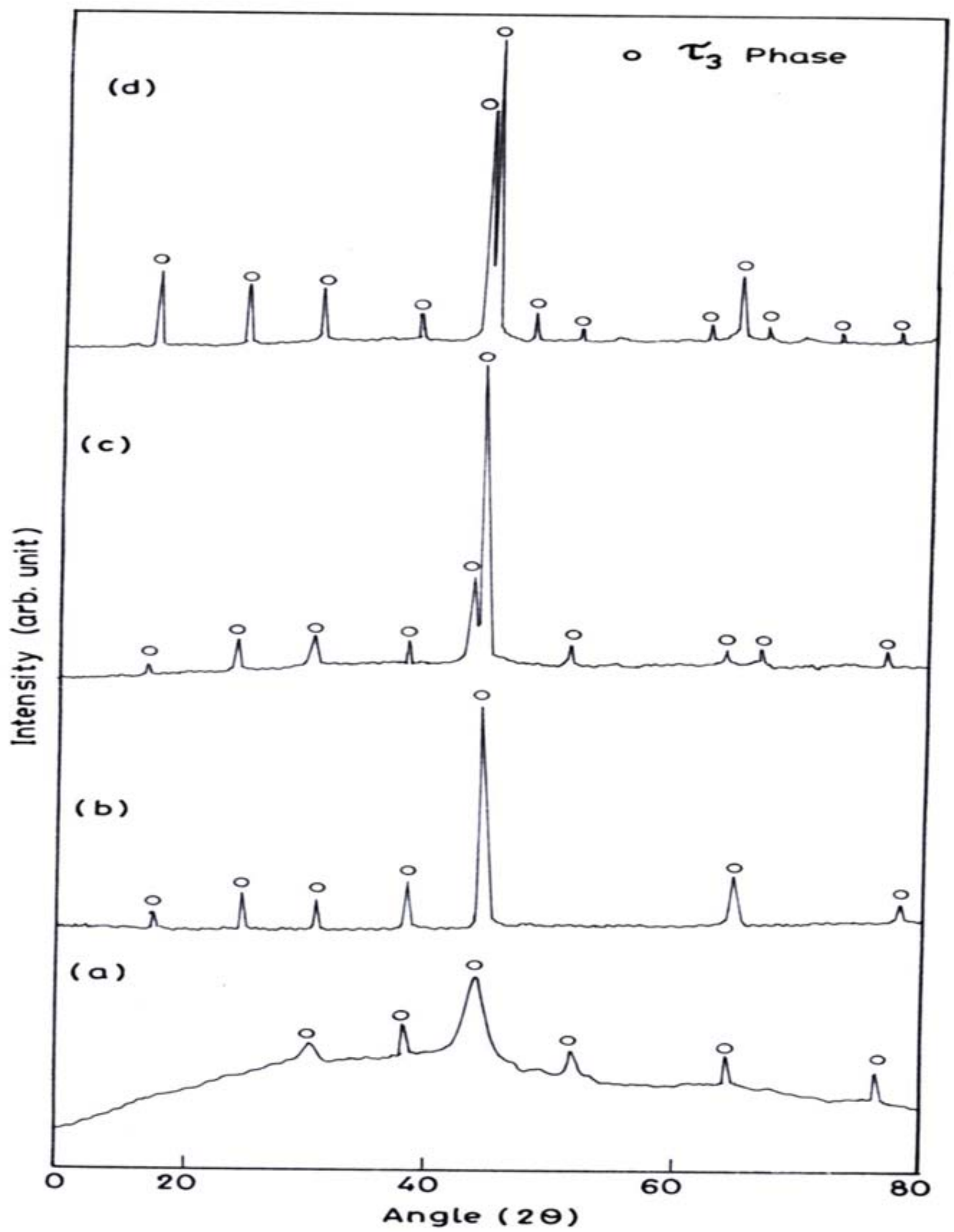

Fig.3: XRD patterns obtained from 100h MM (a) after annealing at 700 °C for 20h-60h respectively (b-d) . In all of the cases the peak broadening is found to be reduced compared with that in the as-milled condition. This can be attributed to strain relaxation along with slight domain coarsening

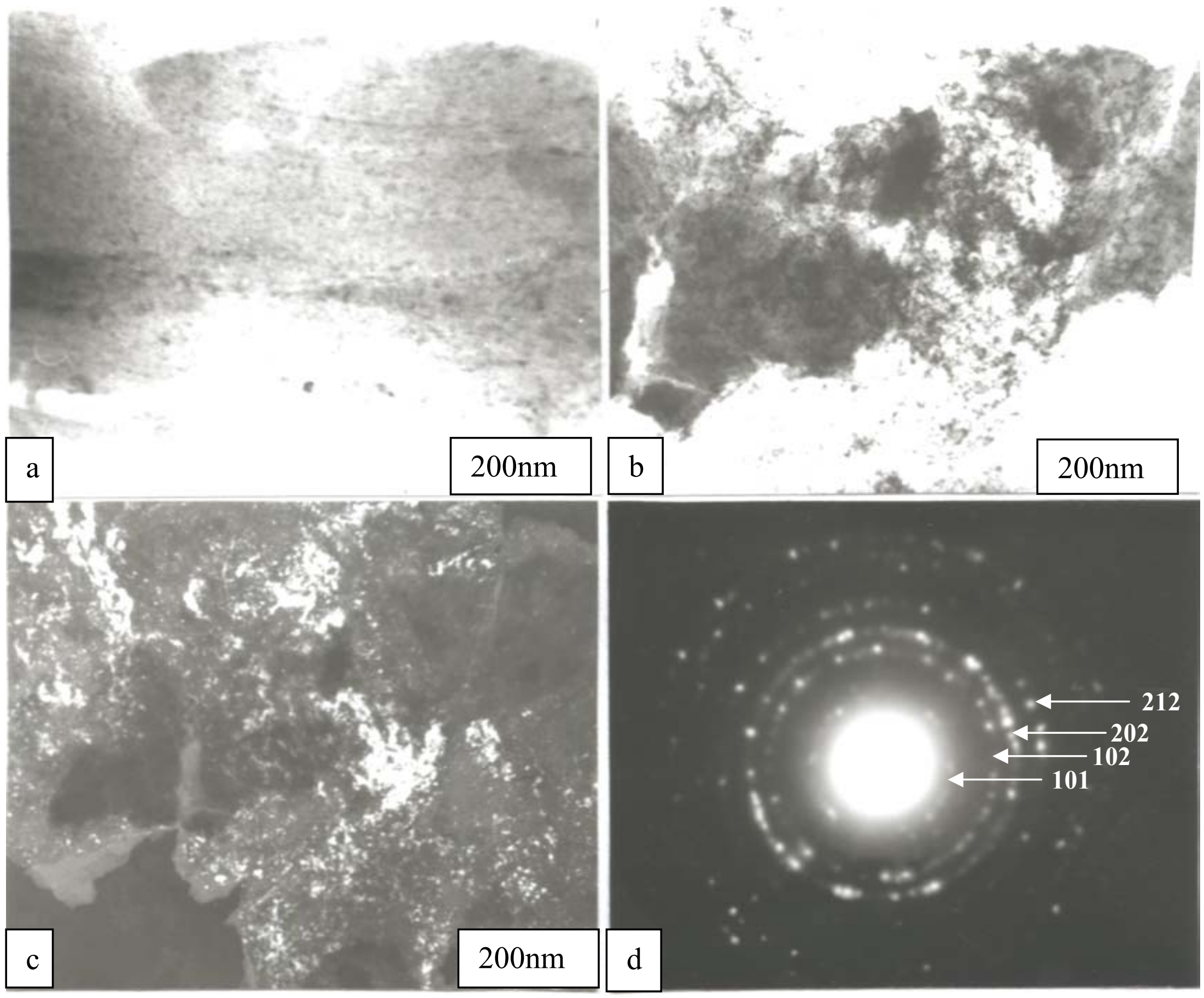

Fig.4: (a) Bright field TEM image of 60h MM, (b) Bright field TEM image of 100 h MM (c ) Dark-field TEM image of powder sample after milling for 100 h, showing aggregates of nanocrystals of the order of 10–20nm in size.(d) The corresponding diffraction pattern shows a ring pattern from the $\tau$3phase. .

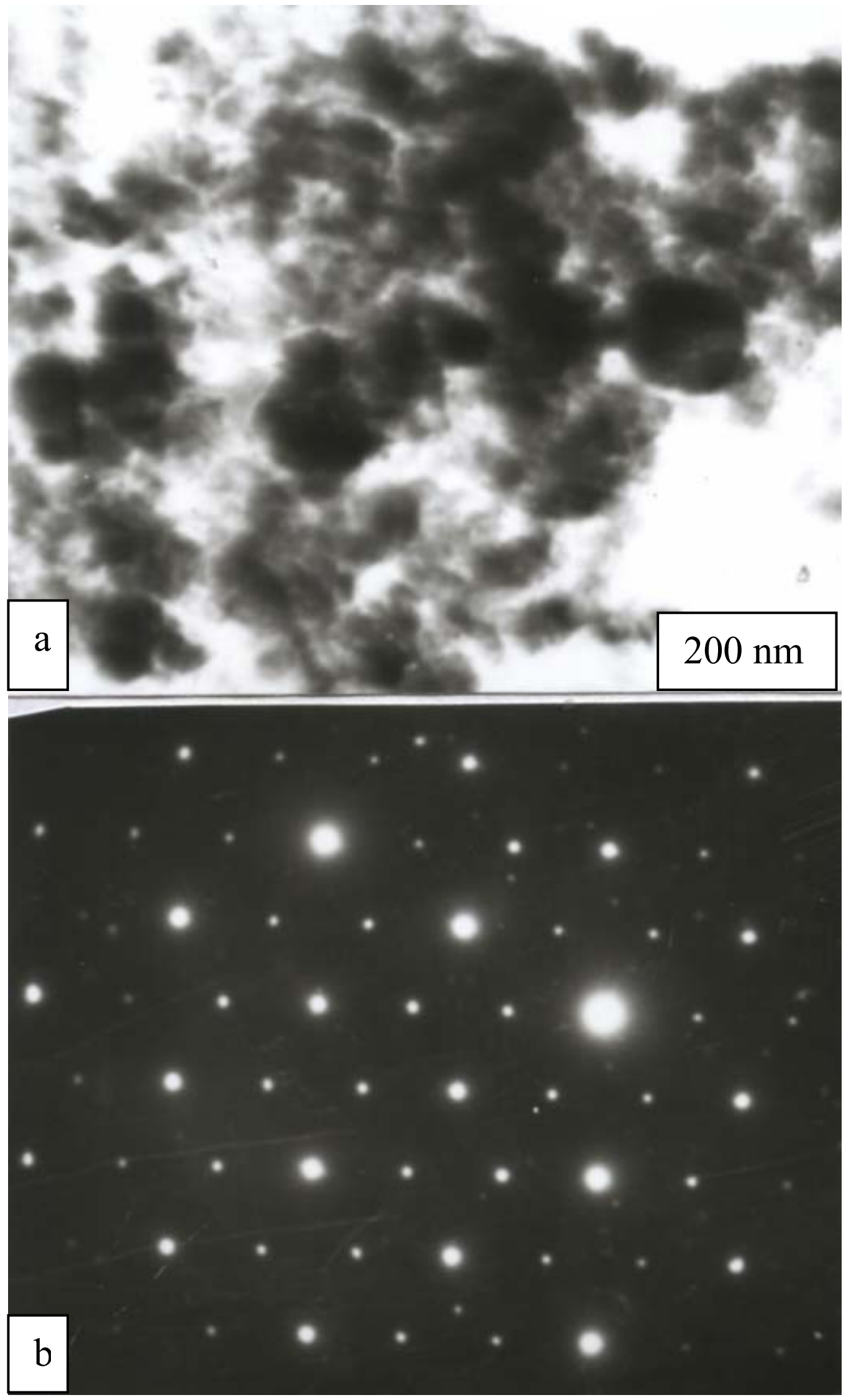

Fig.5: The TEM microstructure of 100h MM and 60 h annealed at 700 ºC powders (b) corresponding diffraction patterns, the indexing of the SADs reveals the presence of τ3 phase